\documentclass[%
 prl,
 reprint,
 amsmath,amssymb,
 aps,
]{revtex4-2}

\usepackage{graphicx}

\begin{document}


\title{Cavity optomechanics in an infinite cylinder}

\author{Samar Deep}
\altaffiliation{These two authors contributed equally to this work.}
\author{Emily Eadie}
\altaffiliation{These two authors contributed equally to this work.}
\author{Pablo Bianucci}
 \email{pablo.bianucci@concordia.ca}
\affiliation{%
 Department of Physics, Concordia University, 7141 Sherbrooke St West, Montreal QC, Canada
}%

\date{May 27, 2026}

\begin{abstract}
We report the observation of optically-induced mechanical oscillations in flame-produced surface nanoscale axial photonic resonators (SNAPRs). 
The frequency of the optically-excited mechanical modes in these SNAPRs is not affected by the particulars of the optical axial confinement, and matches with the theoretically predicted frequency for the fundamental breathing mode of an infinite solid cylinder. When the SNAPR fabrication recipe is changed, changing the optical mode spectrum, the behaviour of the observed mechanical modes remains unchanged. This shows a degree of geometrical uncoupling between the optical and mechanical degrees of freedom, which in turn implies the effective radius variation causing axial optical confinement in our SNAPRs is dominated by a refractive index change rather than a physical radius change.
\end{abstract}

\maketitle


\section{\label{sec:intro}Introduction}

The interaction of light with mechanical vibrations has long been an attractive research topic. For instance, the scattering of light waves with acoustic traveling waves, known as Brillouin-Mandelstam scattering\cite{BrillouinAP1922}, has been of critical importance in many fields\cite{MoritzAPR22}, including optical fiber communications. It is through the study of guided acoustic-wave Brillouin scattering (GAWBS) in optical fibers that the first coupling between light and transverse acoustic modes was found\cite{ShelbyPRB1985}. Since optical fiber can be considered as an infinite glass cylinder, the mechanical vibrations associated with acoustic waves in fibers can be well described by the Pochhammer-Chree equation\cite{PochhammerJRM1876,ChreeTCPS1889}, which allows for a quasi-analytical solution of the mechanical frequency spectrum.

In the context of optical microresonators, there has been much work in the last two decades on cavity optomechanics, the interaction between the light confined inside an optical cavity and mechanical vibrations\cite{AspelmeyerRMP14}. The optomechanics of whispering gallery mode (WGM) resonators has been a particular focus of attention\cite{SchliesserBCh14}, with demonstrations made on varied geometries such as spheres\cite{CarmonPRL2005}, toroids\cite{RokhsariOE05}, bottles\cite{AsanoOE2016,AsanoPRAP2024}, and bubbles\cite{FarnesiSREP17,FarnesiOMX211}. A more recent advance is the understanding that many cavity optomechanics experiments can be considered a particular case of intramodal forward Brillouin scattering\cite{WiederheckerAPLP19}. A common feature of these optomechanical systems is that both the optical and the mechanical modes depend strongly on the cavity morphology. Thus, any change in the resonator geometry will affect both optical and mechanical modes, potentially in complicated ways. In addition, numerical modeling (generally using the finite element method) tends to be required to identify the optical and mechanical modes. 

\begin{figure}
    \centering
    \includegraphics[width=0.75\linewidth]{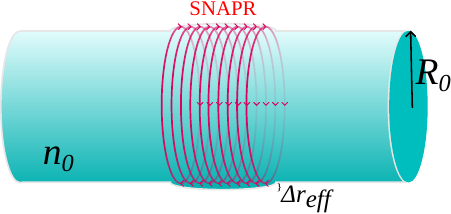}
    \caption{Schematic of a SNAPR and its optical confinement.}
    \label{fig:SNAPRsc}
\end{figure}

Surface nanoscale axial photonic resonators (SNAPRs) are WGM resonators that can be described as optical bottle resonators with a nanoscale effective radius variation, ($r_{\mbox{\scriptsize eff}} = n_0*R_0$, where $n_0$ is the refractive index and $R_0$ is the unperturbed radius) of a dielectric cylinder, as illustrated in Figure \ref{fig:SNAPRsc}. Thanks to the small magnitude of the perturbation, on the order of tens of nanometers, the axial degree of freedom can be approximately separated from the others, which reduces the problem of finding the optical modes to that of solving a Schr\"odinger-like one-dimensional equation\cite{SumetskyOE12}. SNAPRs can have a very clean mode structure, where identification of axial modes (modes that only differ in their axial mode number) is straightforward. SNAPRs have been fabricated using many different methods, including CO$_2$ laser irradiation\cite{SumetskyOL17}, flames\cite{SumetskyLPL2022}, femtosecond lasers\cite{YuOL2019} and fusion splicers\cite{LiuOPL2024}. SNAPRs have been proposed as a platform for the generation of optical frequency combs\cite{KolesnikovaOE2022,EadieOE25}, and low-distortion microscopic optical delay lines\cite{SumetskySREP2015}, among the highest profile applications.

Since the optical behaviour of SNAPRs comes from a variation of its effective radius, it is not possible to clearly separate the contributions of the change in physical radius and refractive index from its optical spectra. If we were to assume that the dominant factor is a localized change in refractive index, leaving the physical radius constant, we would expect an interesting phenomenon: the mechanical modes would continue to behave as those of an infinite cylinder, while still interacting with the confined optical mode. In this work, we show experimental evidence that the mechanical modes in our SNAPRs continue to behave as those of an infinite cylinder, so that resonators with different optical axial profiles still share the same mechanical modes.

\section{\label{sec:tamodes}Transverse mechanical modes in a cylinder}

\begin{figure}
    \centering
    \includegraphics[width=0.75\linewidth]{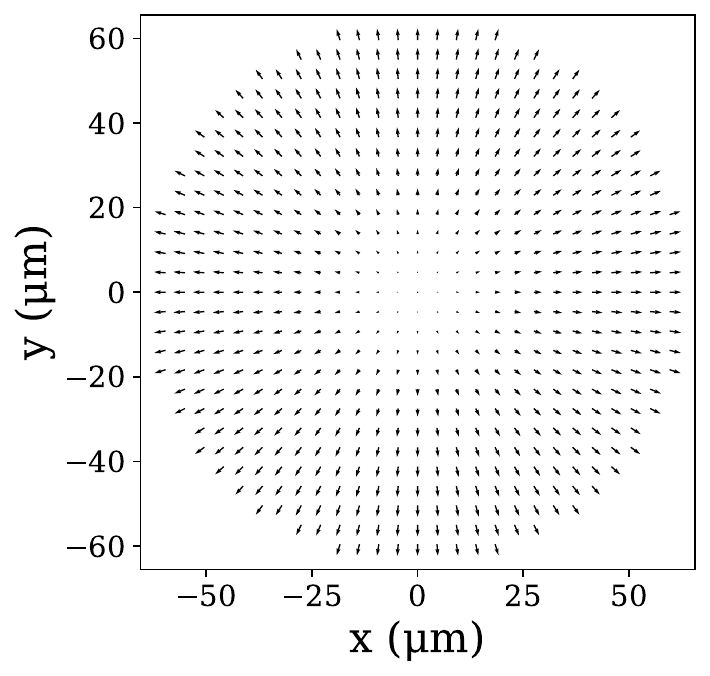}
    \caption{Displacement field of the $R_{01}$ transverse mechanical mode.}
    \label{fig:r01disp}
\end{figure}

The problem of describing the mechanical vibrations of an infinite elastic rod was first approached by Pochhammer\cite{PochhammerJRM1876}, and later independently by Chree\cite{ChreeTCPS1889}, in the final quarter of the 19th century. 
Solutions to this equation describe elastic waves traveling along the cylinder's long axis, and fall into three main families: torsional, longitudinal, and flexural\cite{ViolaBKCh07}. Torsional modes involve purely azimuthal motion, longitudinal modes can display both radial and axial motion, and flexural modes exhibit a complex pattern of displacements. The low-frequency end of the mechanical spectrum is host to modes with a ``breathing" motion, where the cylinder expands and contracts regularly, which belong to the family of longitudinal modes with radial-only displacements. In the context of optical fibers, these modes are known as ``guided acoustic waves''\cite{ShelbyPRB1985}, and denoted as $R_{0m}$, where $m$ is the number of nodes of the displacement field along the radial direction. Using $V_{\mbox{\tiny T}} = 3740$ m/s and $V_{\mbox{\tiny L}} = 5996$ m/s for the transverse and longitudinal sound speeds of silica, respectively, the oscillation frequency of the fundamental $R_{01}$ mode, can be found to be\cite{StillerPHD2011}
\begin{equation}
  \nu_{01} \approx \frac{V_{\mbox{\tiny T}}}{D} \frac{3.205}{\pi},
  \label{eq:mfreq}
\end{equation}
where $D$ is the cylinder's diameter.
A typical displacement field for the $R_{01}$ mode is illustrated in Figure \ref{fig:r01disp}, indicating the ``breathing" nature of the motion.

\section{\label{sec:SNAPRs}Surface Nanoscale Axial Photonic Resonators}

We fabricated SNAPRs by a flame-based thermal treatment\cite{SumetskyLPL2022,HannaAPC2024} of Corning SMF-28e optical fiber, with a nominal diameter of $125 \pm 0.7$ $\mu$m. The flame was produced by an oxygen-hydrogen torch, and the operational parameters consisted of the gap between the torch tip, the gas pressure, and the time of exposure to the flame. We fabricated and measured seven SNAPRs with slightly different parameters (detailed parameters can be found in the Supplementary Information). 

\begin{figure}
    \centering
    \includegraphics[width=0.85\linewidth]{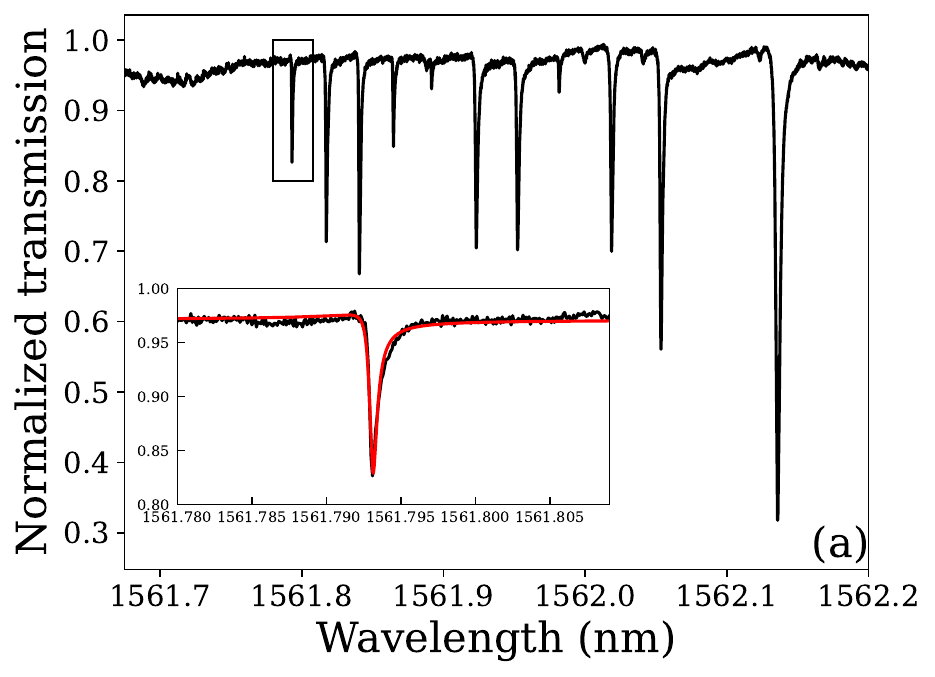}
    \includegraphics[width=0.85\linewidth]{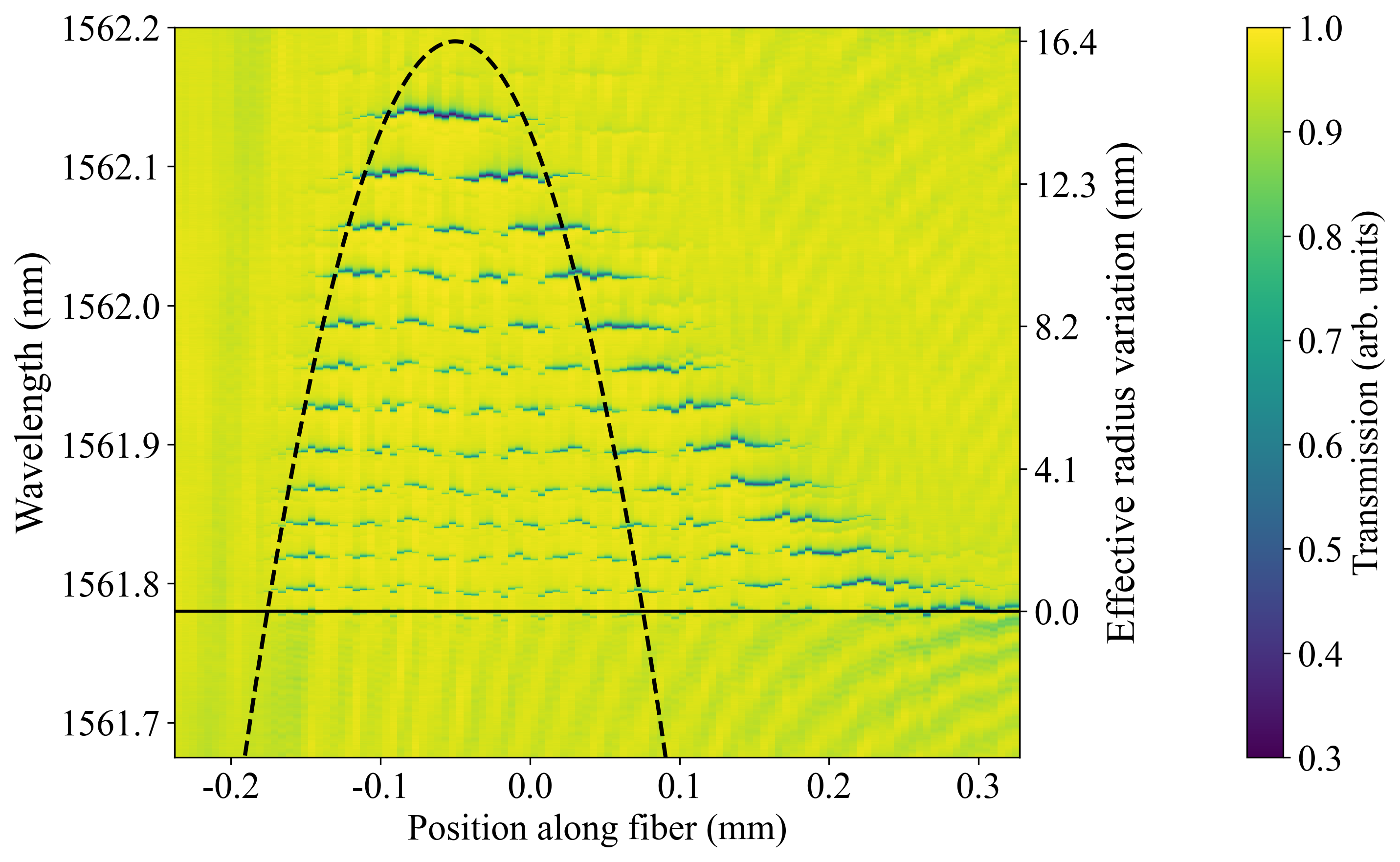}
    \caption{(a) Transmission spectrum of a SNAPR fabricated with a 125 $\mu$m diameter optical fiber, showing resonant modes as dips. Inset: Lorentzian fit of the boxed mode, with a loaded optical Q factor of $2.3\times 10^{6}$. (b) Spectrogram of the same SNAPR, focused on the same group of modes. The spectrogram shows a maximum effective radius variation (left scale) close to 14.2 nm. The black dotted line shows a parabolic fit used to estimate the axial curvature.}
    \label{fig:SNAPR}
\end{figure}

The SNAPRs were characterized using a standard evanescent spectroscopy technique\cite{SumetskyOL10}. The laser wavelength is swept while the transmission is recorded, using a low laser power ($\approx 2 mW$) to stay in the linear regime. Dips in the transmission correspond to resonant optical modes, such as illustrated in Figure \ref{fig:SNAPR}(a). Observed loaded optical Q-factors were in the $10^6-10^7$ range. By displacing the tapered fiber along the SNAPR, we recorded a spectrogram (Figure \ref{fig:SNAPR}(b) shows a representative example) that is related to the spatial distribution of the electric field in the resonant modes. From the spectrogram we can extract the maximum effective radius variation, which for the resonator in \ref{fig:SNAPR}(b) is close to 14.2 nm. The spatial profile is asymmetric away from the top portion of the SNAPR due to the fabrication parameters. Fitting the top part of the SNAPR with a parabolic function, we can estimate the axial radius of curvature for the effective radius variation to be of $R_a \approx -0.52$ m. The seven measured SNAPRs were prepared with slightly different recipes that resulted in different mode distributions (other spectrograms can be found in the Supplementary Information).

\section{\label{sec:om}Optomechanical interactions in SNAPRs}

\begin{figure}
    \centering
    \includegraphics[width=0.85\linewidth]{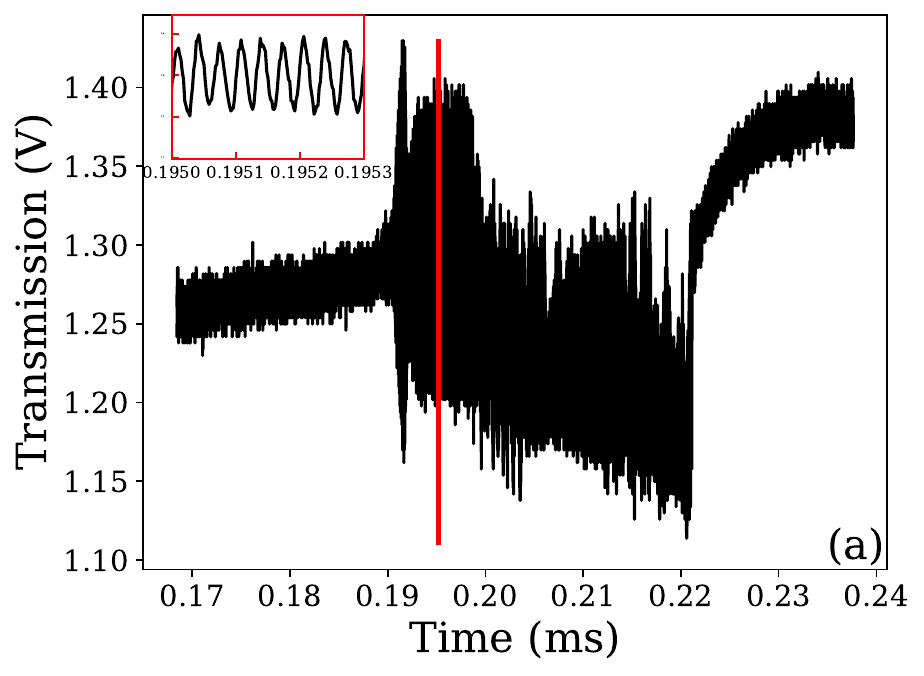}
    \includegraphics[width=0.85\linewidth]{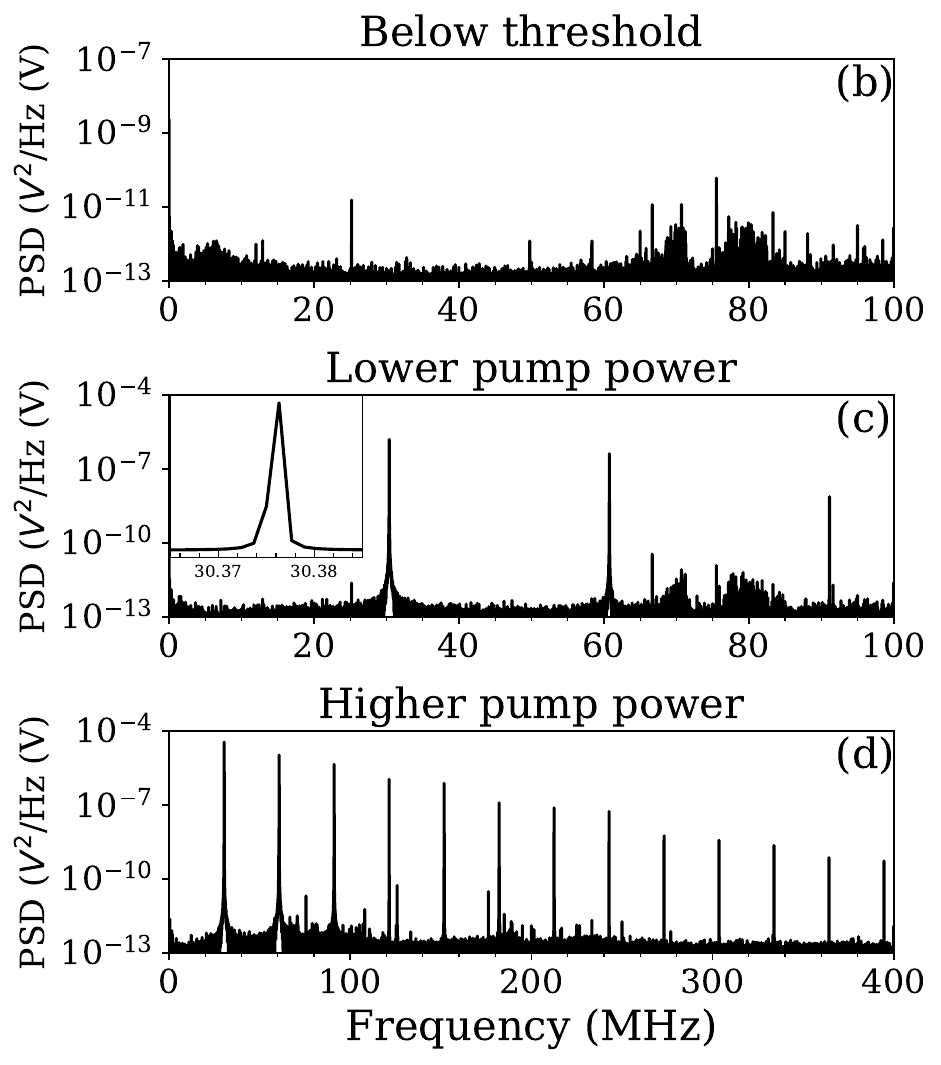}
    \caption{(a) Transmission spectrum showing optomechanical noise. The inset shows a zoom into the transmission showing the oscillatory nature of the noise. (b)-(d) RF spectrum of the transmitted signal at different pump powers: (b) below the optomechanical oscillation threshold; (c) at a lower value above the threshold (200 mW), showing in an inset the fundamental peak near 30 MHz; (d) well above threshold (700 mW), where up to 12 harmonics can be seen.}
    \label{fig:OM}
\end{figure}

When using high excitation powers (above 100 or 200 mW, depending on the resonator), we start seeing broadening of the resonant dips in the optical signal. This broadening is usually attributed to a thermal nonlinearity\cite{CarmonOE2004} and can be used to keep the laser locked to the resonance. As the power is increased, the broadened feature shows a significant oscillatory amplitude noise (see Figure \ref{fig:OM}(a)). When the time signal is observed at a fine scale, the oscillations look quasi-harmonic in nature (as can be appreciated in the inset of Figure \ref{fig:OM}{a}). These oscillations observed while sweeping the pump wavelength have been found to be typical of optomechanical interactions in microresonatos\cite{RokhsariOE05}. For a clear identification of the frequencies present in the optomechanical signal, we pumped the resonator with a constant wavelength, which was manually adjusted to maximize the amplitude of oscillations in the transmission signal. The photodiode signal was fed to an electrical spectrum analyzer, which measured its power spectrum. Below the optomechanical oscillations threshold, no obvious features appear, just some likely noise with very low amplitudes (Figure \ref{fig:OM}(b)). Above the threshold, but at the lower end of the pump power range (200 mW, Figure \ref{fig:OM}(c)), we see the fundamental frequency near 30 MHz, with three harmonics present. At the higher end (700 mW, Figure \ref{fig:OM}(d)), we still see the fundamental frequency, but with a much larger set of visible harmonics (12, limited by our collection equipment bandwidth). The Q factor of the mechanical mode can be determined by the full-width half-maximum of the electrical spectrum peak when measured as close as possible to the threshold. Even at the lowest excitation pump powers showing optomechanical oscillations, the width of the peak is too narrow to resolve properly with our 1.315 kHz frequency resolution. We can still use this fact to establish a lower bound to the mechanical quality factor, $Q_{\mbox{m}} \geq 30.4\ \mbox{MHz}/1.315\ \mbox{kHz} \approx 23,100$.

\begin{figure}
    \centering
    \includegraphics[width=0.85\linewidth]{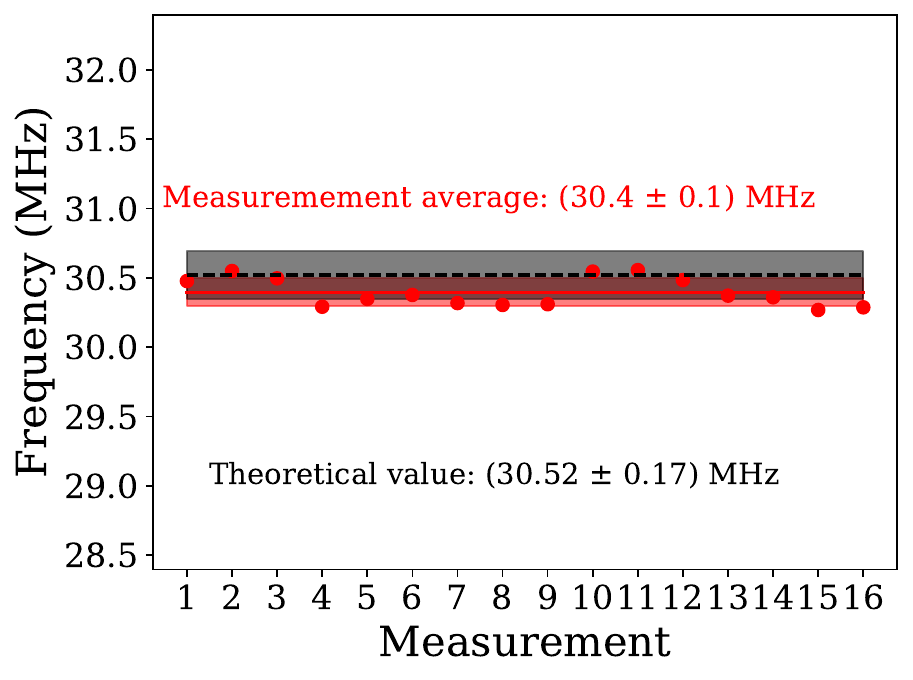}
    \caption{Mechanical frequency of the lowest mechanical mode for measurements in different positions of different SNAPRs. The red shaded range shows the uncertainty range for the experimental measurements, while the gray shaded range shows the theoretically computed range.}
    \label{fig:125um}
\end{figure}

In all seven SNAPRs, we observed optomechanical resonances at different positions of the coupling fiber, corresponding to different axial optical modes. Altogether, this provided sixteen different measurements taken at excitation powers between 200 mW and 700 mW, measured at the output of the fiber amplifier. In each case, we extracted the frequency of the fundamental mechanical vibration from the spectrum analyzer data. The results are shown in Figure \ref{fig:125um}. All the measured mechanical frequencies are close to each other, with an average frequency value of $(30.4 \pm 0.1)$ MHz. Since the measurements were made on different optical modes of non-identical SNAPRs, the agreement of the resonant frequencies (on the order of 0.3\%) is quite remarkable. This strongly suggests that the particulars of the axial optical confinement (that is, the specific effective radius profile) does not significantly affect the mechanical vibration frequency. 

If we consider our SNAPR as an infinite silica cylinder, with a diameter of $(125 \pm 0.7)$ $\mu$m, we can use equation \ref{eq:mfreq}, to find the frequency range for the fundamental breathing mode $\nu_{01} = (30.52 \pm 0.17)$ MHz. As it can be seen in Figure \ref{fig:125um}, the two ranges overlap, giving us more confidence that the optically excited mechanical mode in the SNAPRs corresponds to the $R_{01}$ fundamental breathing mode.

\begin{figure}
    \centering
    \includegraphics[width=0.85\linewidth]{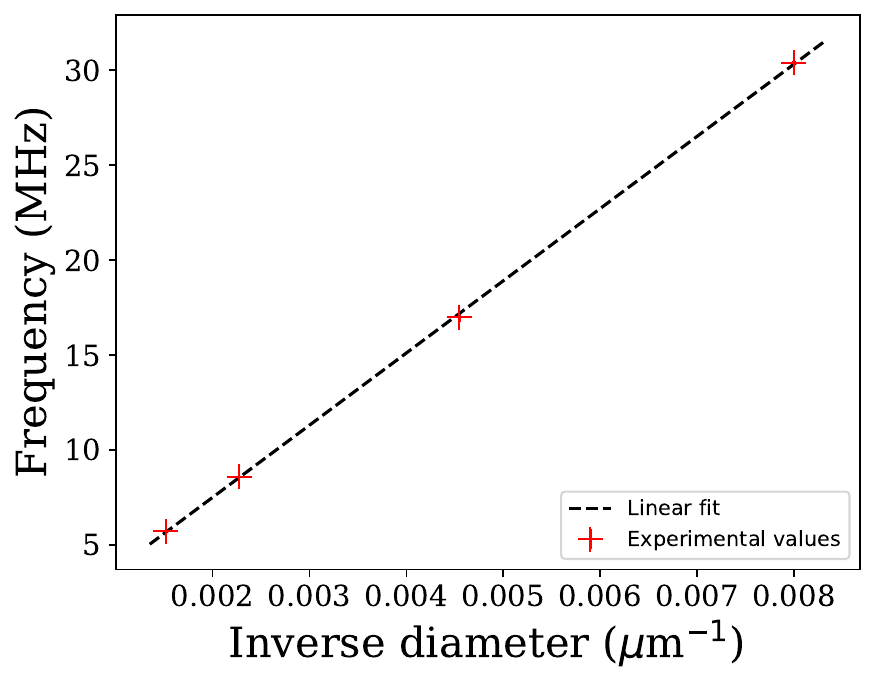}
    \caption{Mechanical frequency of the breathing mechanical mode for SNAPRs made with fibers of different radii. The crosses correspond to the experimentally found frequencies, while the dotted line is a fit with Equation \ref{eq:mfreq}}.  
    \label{fig:radii}
\end{figure}

We performed similar measurements in SNAPRs fabricated out of optical fibers with other nominal diameters (220 $\mu$m, 440 $\mu$m, and 660 $\mu$m) to confirm whether the predictions from equation \ref{eq:mfreq} would match the observed mechanical frequencies. When plotted as a function of the inverse fiber diameter, the results form the expected straight line as can be seen in Figure \ref{fig:radii}. Furthermore, from a linear regression we can find the slope and, inverting Equation \ref{eq:mfreq}, a value for $V_{\mbox{\tiny T}}= (3730 \pm 29)$ m/s. The experimental data is well described, both qualitatively and quantitatively, by equation \ref{eq:mfreq}.

The robustness of the mode's mechanical frequency, combined with the good fit to the infinite cylinder model for varying SNAPR radius makes it clear that the optical confinement is only weakly (if at all) affecting the mechanical motion. Unlike the case of most monolithic resonators, where both the optical and mechanical modes are strongly dependent on the cavity geometry, in SNAPRs the mechanical and optical degrees of freedom are coupled only in the transverse plane. Axial coupling is negligible because the details of optical axial confinement have little effect on the mechanical frequency.

The axial uncoupling allows us to shed light on the origin of the effective radius variation that creates the SNAPR. Since the radius and refractive index contribute equally to the effective radius, our optical data alone cannot distinguish their individual contributions. However, information from the mechanical vibrations provides an important clue. If the effective radius change were dominated by a physical radius variation, we would expect confined mechanical modes\cite{SumetskyOL17}. For a silica SNAPR like the one in Figure \ref{fig:SNAPR}, these modes would have a mechanical free spectral of roughly 17 KHz. Given our approximate frequency resolution of 1.315 KHz, such confined modes should be observable in our mechanical spectra. As seen in Figure \ref{fig:OM}(b), however, we have a clean spectrum composed solely of the fundamental breathing mode and its harmonics . This strongly implies that the modification that created our SNAPRs is dominated by a refractive index change rather than a physical radius modification. This is the first time a specific mechanism has been attributed to the effective radius variation in a SNAPR.

\section{\label{sec:concl}Conclusion}

At high optical pump powers, we have observed optomechanical interactions in SNAPRs. The spectra of the optical transmission show optomechanical oscillations with a single dominant frequency and its harmonics. The frequency of this mechanical mode matches well that of the transverse $R_{01}$ mode expected in an infinite cylinder, independently of the particulars of the optical axial confinement. Mechanical modes observed in SNAPRs with different diameters also match well the expected oscillation frequencies, further confirming the identification of the mechanical oscillations with the $R_{01}$ mode. Finally, the fact that the axial optical confinement does not seem to affect the mechanical modes shows that, in our flame-fabricated SNAPRs, the dominant mechanism for the effective radius variation is a change in refractive index rather than a change in physical radius. This knowledge would be useful in conceiving future experiments probing other mechanical modes in cylinders. We can also see potential applications in the design of optomechanical devices with robust mechanical frequencies, such as fiber phase modulators.


\begin{acknowledgments}
We would like to acknowledge funding from the Natural Sciences and Engineering Research Council (NSERC) of Canada, individual Discovery grant (RGPIN-2019-06988), and from the Centre d'Optique, Photonique et laser (COPL, 2025-RSMA-340950), a strategic cluster from the Fonds de Recherche du Qu\'{e}bec--Nature et Technologies (FRQ-NT). E.E. acknowledges support from the FRQ-NT through a Doctoral Research Scholarship (award no. 2002241) and from NSERC through a Canada Graduate Research Scholarship – Doctoral (CGRS D, application no. 612419).
\end{acknowledgments}

\begin{acknowledgments}
The data that support the findings of this article are not publicly available. The data are available from the authors upon reasonable request.
\end{acknowledgments}


\bibliography{OMCyl}

\end{document}


\textbf{\LARGE Supplementary information}

\section{SNAPR fabrication parameters}

\begin{figure}[ht]
    \centering
    \includegraphics[width=0.475\textwidth]{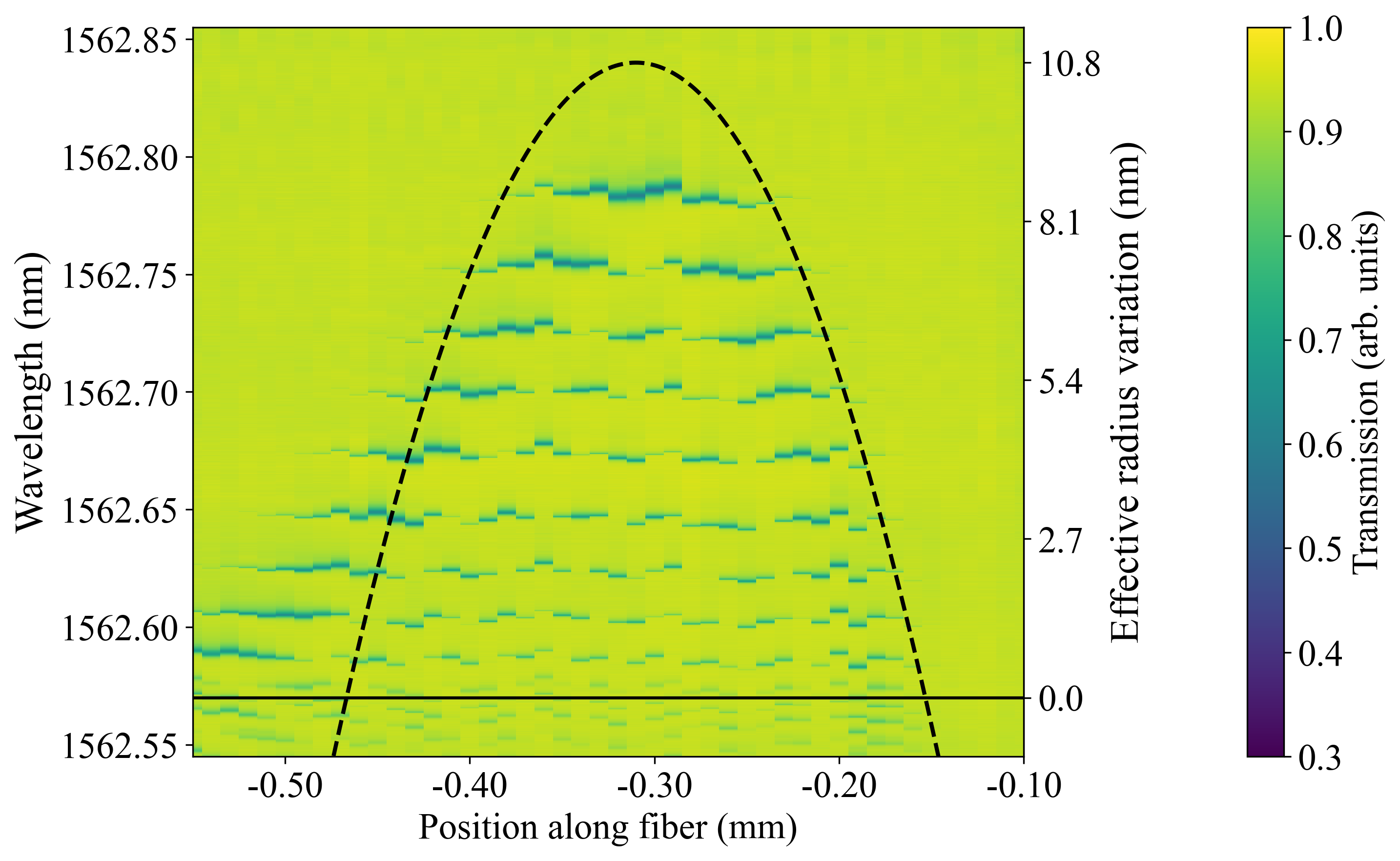}
    \includegraphics[width=0.475\textwidth]{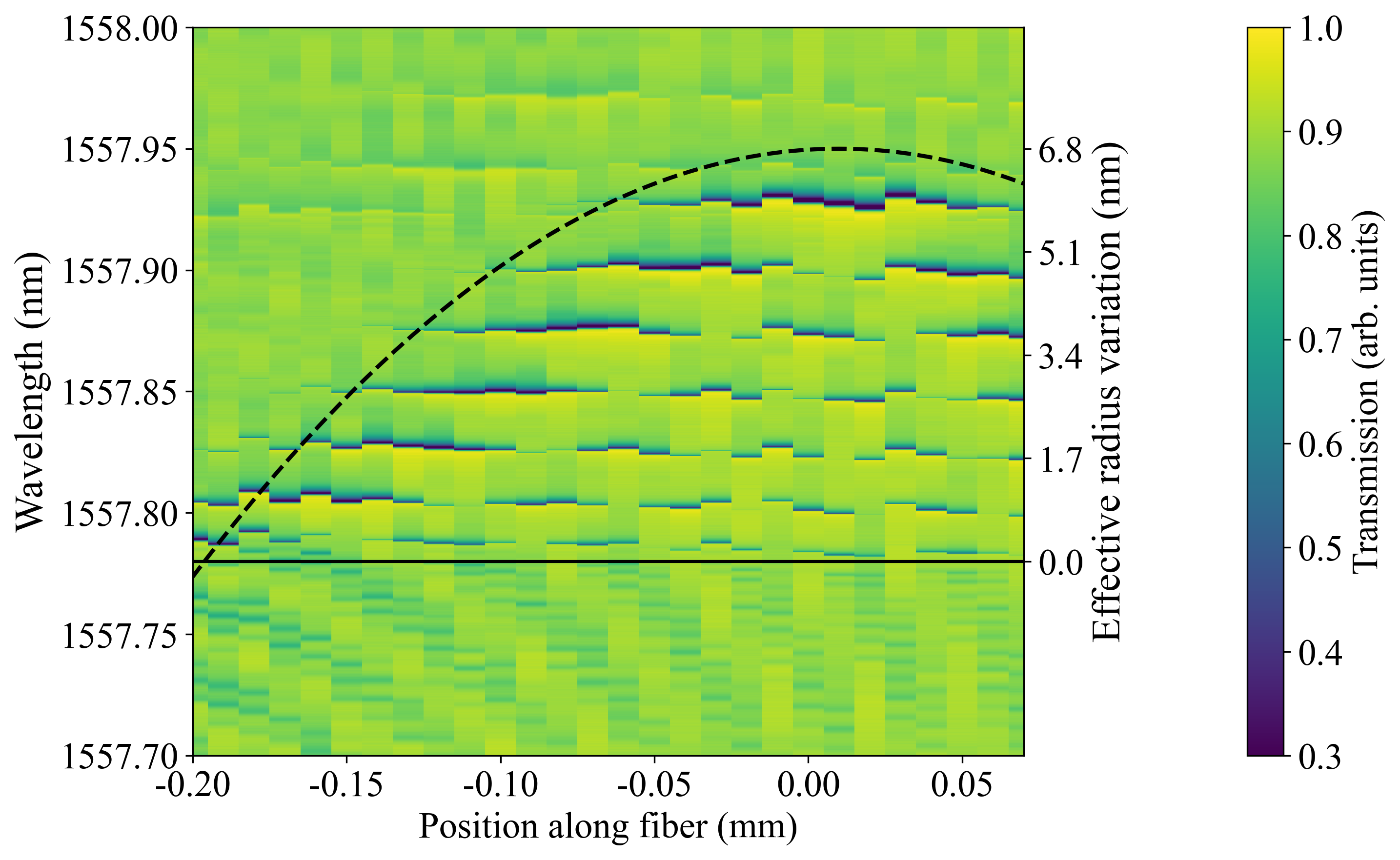}
    \caption{Spectrograms for SNAPRs \#6 (left) and \#7 (right). It can be seen that they have different axial curvatures and maximum effective radius variations, evidenced by the different number of axial modes present. They are also different from SNAPR \#3, shown in the main text.}
    \label{fig:spgs}
\end{figure}

Our SNAPRs are fabricated by exposing the fiber to the flame of an oxy-hydrogen torch. There are three experimental parameters under our control that modify the resulting SNAPR: The torch-fiber gap, the exposure time, and the pressure differencial between the gas and atmospheric pressure (referred to as ``torch pressure" in the rest of the document). As a general rule, smaller gaps, longer exposure times, and higher pressures lead to ``deeper" SNAPRs, with a higher effectiver radius variation and more axial modes. Table \ref{tab:params} shows the parameters for each one of the SNAPRs fabricated out of regular telecom fiber (125 $\mu$m diameter). Figure \ref{fig:spgs}
shows spectrograms for two of those SNAPRs.

\begin{table}[ht]
    \centering
    \begin{tabular}{|c|c|c|c|} \hline
      SNAPR number   & Torch-fiber gap (mm) & Exposure time (s) & Torch pressure (PSI)  \\ \hline
         1 &  4 & 5 & 5.3 \\ \hline
         2 &  4 & 4 & 5.3 \\ \hline
         3 &  3 & 4 & 5.58 \\ \hline
         4 &  5 & 4 & 5.29 \\ \hline
         5 &  4 & 3 & 5.39 \\ \hline
         6 &  4 & 4 & 4.9 \\ \hline
         7 &  4 & 4 & 5.8 \\ \hline
    \end{tabular}
    \caption{Fabrication parameters for the 125-$\mu$m-diameter SNAPRs.}
    \label{tab:params}
\end{table}

\section{Experimental setup}

\begin{figure}
    \centering
      \begin{tikzpicture}[
        node distance=1.25cm,
        instrument/.style={rectangle, draw, minimum width=2.5cm, minimum height=1cm, text centered},
        edfa/.style={isosceles triangle, draw, minimum width=2.0cm, minimum height=1cm, text centered},
        snapr_box/.style={cylinder, shape border rotate=90, draw, minimum width=0.05cm, minimum height=2.0cm, text centered},
        snapr_label/.style={below=of #1, text centered}
        arrow/.style={->, thick}
                         ]
        \node (tls) [instrument, align=center] {Tunable\\Laser Source};
        \node (isolator) [instrument, below=of tls, align=center] {Optical\\isolator};
        \node (edfa) [edfa, right=of isolator] {EDFA};
        \node (PC) [instrument, right=of edfa, align=center] {Polarization\\controller};
        \node (left_taper) [right=1 cm of PC] {};
        \node (snapr_box) [snapr_box, right=0.5cm of left_taper] {};
        \node (right_taper) [right=0.5 cm of snapr_box.center] {};
        \node (snapr_label) [below=0.1 cm of snapr_box] {SNAPR};
        \node (photodiode) [instrument, right=1 cm of right_taper] {Photodiode};
        \node (oscilloscope) [instrument, above=of photodiode] {Oscilloscope};
        \node (esa) [instrument, below=of photodiode, align=center] {Electrical\\Spectrum\\Analyzer};

        \draw [line width=2pt] (tls) -- (isolator);
        \draw [line width=2pt] (isolator) -- (edfa);
        \draw [line width=2pt] (edfa.apex) -- (PC);
        \draw [line width=2pt] (PC) -- (left_taper.west);
        \draw [very thin] (left_taper.west) -- (right_taper.east);
        \draw [line width=2pt] (right_taper.east) -- (photodiode);
        \draw [line width=2pt] (photodiode) -- (oscilloscope);
        \draw [line width=2pt](photodiode) -- (esa);
    \end{tikzpicture}
    
    \caption{Schematic of the experimental setup.}
    \label{fig:setup}
\end{figure}
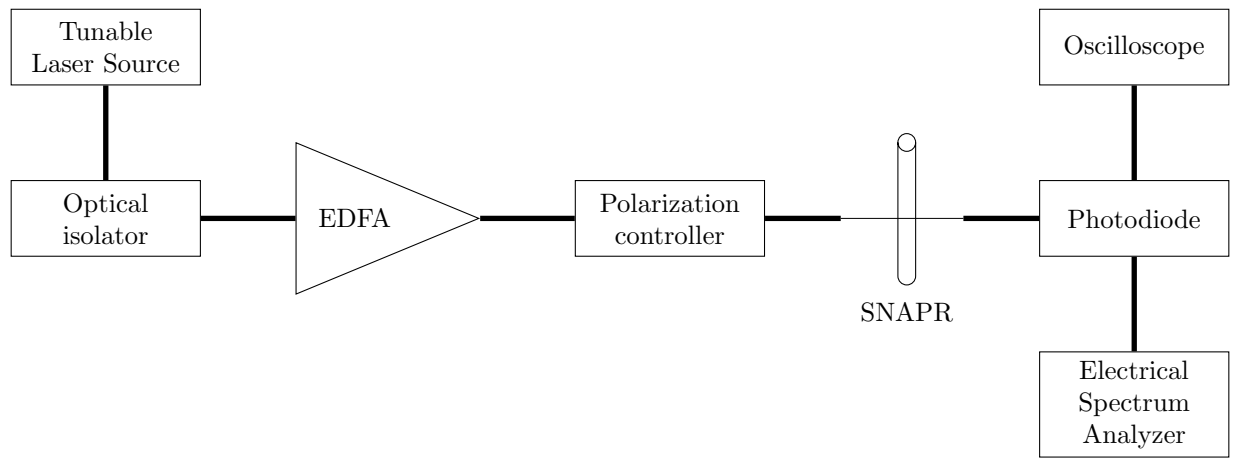

The experiments were performed with a standard evanescent spectroscopy setup, as illustrated in Figure \ref{fig:setup}. A tunable laser source is coupled to an erbium-doped fiber amplifier (EDFA) through an optical isolator. After the EDFA, a polarization controller is used to optimize coupling to a specific polarization family of modes and the resulting light is coupled to the SNAPR through the tapered fiber. The transmission is measured with an amplified photodiode, and the resulting signal is analyzed with both an oscilloscope and electrical spectrum amplifier.
